\documentclass[a4paper,12pt]{article}
\usepackage{amsmath}
\usepackage{epsfig,subfigure}
\usepackage{cite}
\date{}
\begin{document}
\title{
\Large{
\textbf{Sensitive methods for estimating the anchoring strength of
nematic liquid crystals on Langmuir-Blodgett monolayers of fatty 
acids}
}
}
\author{
Valentina S. U. Fazio, Francesca Nannelli, Lachezar Komitov \\
\textit{\small{
Department of Microelectronics and Nanoscience,
Liquid Crystal Physics,
}}\\
\textit{\small{
Chalmers University of Technology \&
G\"oteborg University, SE-41296 G\"oteborg, Sweden
}} \\
}
\normalsize
\normalsize
\maketitle
\begin{abstract}
\noindent
The anchoring of the nematic liquid crystal  
N-(p-methoxybenzylidene)-p-butylaniline (MBBA) on Langmuir-Blodgett 
monolayers of fatty acids (COOHC$_{n}$H$_{2n+1}$)  
was studied as a function of the length of the fatty acid
alkyl chains, $n$ ($n = 15, 17, 19, 21$).
The monolayers were deposited onto ITO-coated glass plates which were 
used to assemble sandwich cells of various thickness that were 
filled with MBBA in the nematic phase.
The mechanism of relaxation from the flow-induced quasi-planar 
to the surface-induced homeotropic alignment was studied for the four 
aligning monolayers.
It was found that the speed of the relaxation decreases linearly with 
increasing the length of the alkyl chains $n$ which suggests that the 
Langmuir-Blodgett film plays a role in the phenomenon.
This fact was confirmed by a sensitive estimation of the anchoring 
strength of MBBA on the fatty acid monolayers after anchoring breaking 
which takes place at the transition between two electric-field--induced 
turbulent states, denoted as DSM1 and DSM2.
It was found that the threshold electric field for the anchoring 
breaking, which can be considered as a measure of the anchoring 
strength, also decreases linearly as $n$ increases.
Both methods thus possess a high sensitivity in resolving small 
differences in anchoring strength.
In cells coated with mixed Langmuir-Blodgett monolayers of two fatty acids 
($n=15$ and $n=17$) a maximum of the relaxation speed was observed
when the two acids were present in equal amount.
This observation suggests an efficient method for controlling the 
anchoring strength in homeotropic cells by changing the ratio between 
the components of the surfactant film.
\end{abstract}
%


%
\section{Introduction}
For the operation of liquid crystal displays and devices (LCDs) the 
alignment of the liquid crystal (LC) plays a vital role.
In the absence of an external electric or magnetic field (field-free 
condition) the liquid crystal alignment is entirely dictated by the 
LC/substrate interactions.
These interactions have been a subject of intensive study for more 
than two decades with surface anchoring being one of the highlights 
of the surface physics of liquid crystals.
Although the LC/substrate interactions are to a large extend 
understood, the lack of sensitive and reliable techniques for 
measuring the anchoring strength is an obstacle for a fast progress of 
our knowledge in the field.

There are various techniques for measuring the anchoring 
strength\cite{BlinovChigrinov, Sonin, Yokohama-inc}.
In general, they can be divided in two major groups depending on 
whether the measurements have been carried out in a filed-free 
condition or not.
One of the most frequently and widely used method is the Frederiks 
transition\cite{Sonin}. 
Although it gives satisfactory results, in some cases it is not 
sensitive enough to resolve small differences in anchoring strength.

The effort in the development of liquid crystal displays has been 
focussed, generally speaking, on improving the image contrast, the 
resolution, the number of gray levels as well as on increasing the 
size of the display and shortening the response time.
Recently \cite{KomBabMat95, KoiKatsas97}, 
the so called vertical mode (VA) display has begun to attract 
the interest of researchers and engineers mainly for two reasons.
First, the homeotropic (vertical) alignment of the liquid crystal 
required by the VA mode can be easily controlled by an electric field,
due to the weak anchoring conditions.
Second, the contrast is very high and the response time is short.
To realize the VA mode, however, it is necessary to align the liquid 
crystal homeotropically, i.e. with the nematic director
perpendicular to the confining substrates.
Also, the homeotropic anchoring strength is an important parameter
that influences the device performances.

There are a number of methods for obtaining homeotropic alignment
\cite{Jerome91}.
Among them, the most used one is that of treating the substrate 
with a surfactant.
It is known that very densely packed surfactant layers do not give 
homeotropic alignment of good quality.
A looser packing of the surfactant molecules usually gives good 
homeotropic alignment since it allows a certain degree of penetration 
(interdigitation) of the liquid crystal molecules in the surfactant 
layer \cite{HilSte81, HilSte84}.
Homeotropic alignment obtained with different agents differs not only 
by the degree of uniformity, but also by the magnitude of the 
strength of the liquid crystal anchoring, which varies between 
10$^{-5}$ and 10$^{-6}$ J\,m$^{-2}$ (weak anchoring).


In this work the anchoring of MBBA 
(N-(p-methoxybenzylidene)-p-butylaniline)
on Langmuir-Blodgett (LB) monolayers of fatty acids (COOHC$_{n}$H$_{2n+1}$) 
is studied as a function of the length $n$ ($n$ = 15, 17, 19, 21)
of the fatty acid alkyl chains.
Generally, the nematic liquid crystal anchoring is characterized by two 
parameters, $w_{\theta}$ and $w_{\varphi}$, the polar and the azimuthal 
anchoring strength, respectively.
Here only $w_{\theta}$ will be considered and we will refer to it as 
$w$.

In previous studies \cite{FazKomLagMot00} we have shown 
that the liquid crystal nematic capillary flow has a strong impact on 
the surfactant LB monolayer.
The LB molecules are oriented in the azimuthal direction of the flow
with a large pretilt which results in a quasi-planar NLC alignment
with splay-bend deformation and preferred orientation along the flow 
direction\cite{FazKomLag98a, FazKomLagMot00, FazioValentina00}.
Once the flow ceases the LB film relaxes to the equilibrium orientation
in which the LB molecules present a tilt (about 40 degrees) isotropically 
distributed in the plane of the cell which induces homeotropic alignment 
in the NLC bulk.
Here we will show that the dynamics of this process depend on the 
length of the surfactant molecules. 
Also, in \cite{FazKom99} we have shown that above a certain electric 
field threshold a turbulence-to-turbulence (DSM1-DSM2) transition occurs 
in MBBA, which is related to breaking of the anchoring.
The threshold was found to be proportional to the anchoring strength, 
(about 10$^{-6}$ J\,m$^{-2}$) being thus a measure of it.
Here we will show that this turbulence-to-turbulence transition is 
very sensitive to $w$, and it can resolve even the very small 
differences due to the use of slightly different surfactants.
Since homeotropic alignment is generally weak these differences can 
be extremely small.
In this context it is important to underline that the Fredericks 
transition method, the most widely used method for evaluating  
$w$, does not possess the same sensitivity.

\section{Experiment and results}
\subsection{Film preparation}
\begin{figure}
\begin{center}
\epsfig{file=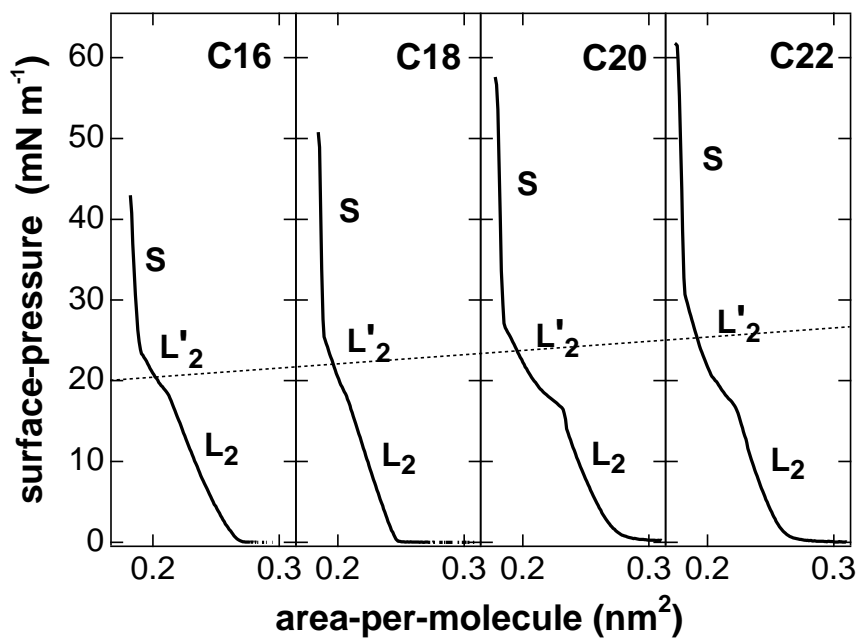, width=0.9\textwidth}
\caption{
Surface-pressure versus area-per-molecule isotherms (20$^{\circ}$C) 
of the fatty acids listed in Table \protect\ref{acids}.
We can distinguish the liquid (\textbf{\textsf{L$_{2}$}}), the liquid-condensed 
(\textbf{\textsf{L$^{\prime}_{2}$}}), and the solid (\textbf{\textsf{S}}) phases 
(see Table \protect\ref{phases} for phase characteristics).
The dashed line indicates the deposition conditions: phase 
\textbf{\textsf{L$^{\prime}_{2}$}} and same area-per-molecule 
(0.2\,nm$^{2}$).
\label{isot_moena00_art}
}
\end{center}
\end{figure}

The long chain fatty acids used in this experiment are listed 
in Table \ref{acids}.
\begin{table}[b]
\caption{
Long chain fatty acids used in this experiment.
\label{acids}
}
\begin{center}
\begin{tabular}{ccc}
\hline
\hline
Common name & Formula & Abbreviation \\
\hline
Palmitic & C$_{15}$H$_{31}$COOH & C16 \\
Stearic & C$_{17}$H$_{35}$COOH & C18 \\
Arachidic & C$_{19}$H$_{39}$COOH & C20 \\
Behenic & C$_{21}$H$_{43}$COOH & C22 \\
\hline
\hline
\end{tabular}
\end{center}
\end{table}
They were spread from chloroform solutions on the surface of ultrapure 
Milli-Q water in a LB trough held in a cleanroom.
The monolayers at the air/water interface were characterized by 
surface-pressure compression isotherms, which are shown in Figure
\ref{isot_moena00_art}.
Three condensed phases were observed for all substances, whose
characteristics are listed in Table \ref{phases}.
\begin{table}
\caption{
Condensed monolayer phases for fatty acids (after Petty 
\protect\cite{Petty}).
\label{phases}
}
\begin{center}
\begin{tabular}{lll}
\hline
\hline
Phase & Name & Characteristics \\
\hline
\textbf{\textsf{L$_{2}$}} & liquid-condensed & tilted molecules \\
\textbf{\textsf{L$^{\prime}_{2}$}} & liquid-condensed & almost upright molecules, \\
& & same compressibility to \textbf{\textsf{L$_{2}$}}\\
\textbf{\textsf{S}} & solid & upright molecules; high collapse pressure; \\
& & less compressible than \textbf{\textsf{L$_{2}$}} and 
\textbf{\textsf{L$^{\prime}_{2}$} }\\
\hline
\hline
\end{tabular}
\end{center}
\end{table}
We chose to deposit the LB films in the \textbf{\textsf{L$^{\prime}_{2}$}}
phase, where the molecules are almost upright and still the monolayers 
are far from collapse \cite{Petty}.
Also, the monolayers were deposited at the same area-per-molecule 
(0.2\,nm$^{2}$), in order to have the same surface density of molecules 
on the plates.

\begin{figure}
\begin{center}
\epsfig{file=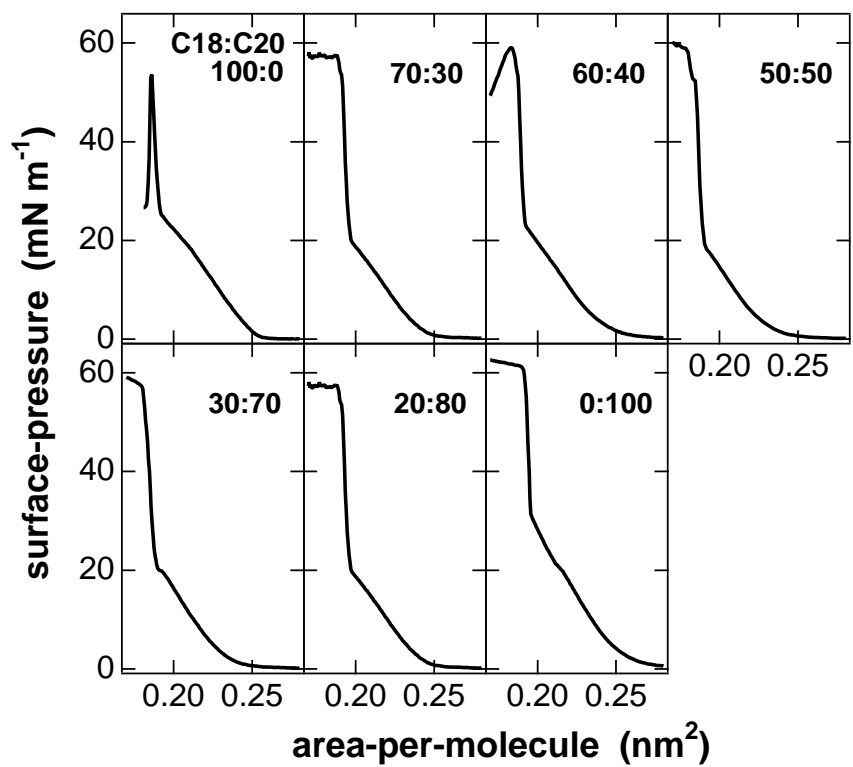, width=0.9\textwidth}
\caption{
\label{mixedisot}
Surface-pressure versus area-per-molecule isotherms (20$^{\circ}$C)
of the C18:C22 mixed monolayers.
The proportions of the two compounds are specified.
}
\end{center}
\end{figure}

We also deposited mixed monolayers of C18 and C22 in different 
proportions.
The miscibility of these two fatty acids at the air/water interface 
has been studied elsewhere\cite{FazKomLag98b}.
The isotherms are shown in Figure \ref{mixedisot}.
The mixed monolayers were deposited at the fixed surface-pressure of 
20\,mN\,m$^{-1}$.
The transfer ratio was always one.

We used ITO coated glass plates, previously carefully cleaned in a 
clean room environment.

\subsection{Alignment dynamics}
\begin{figure}
\begin{center}
\epsfig{file=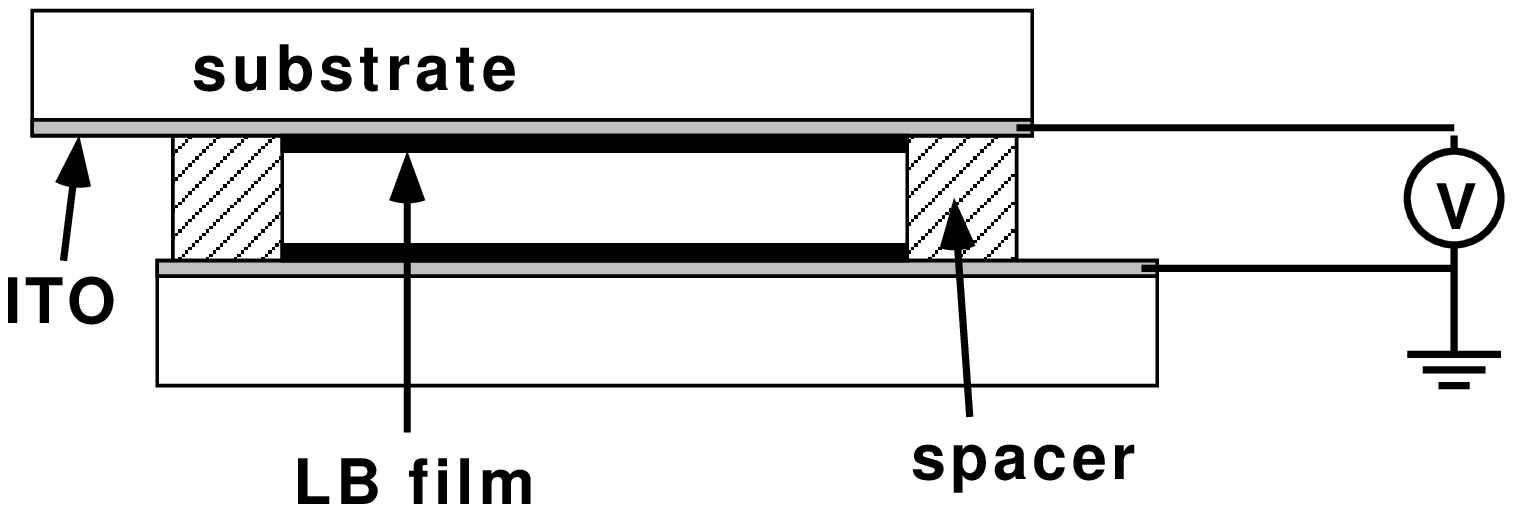, width=0.6\textwidth}
\caption{Cross section of a cell.
\label{cell}
}
\end{center}
\end{figure}

The LB coated plates were used to assemble sandwich cells 
(Figure \ref{cell}) of various thickness.
The cells were filled with MBBA in the nematic phase (room 
temperature).
During filling the NLC adopts a bend-splay--deformed 
quasi-planar alignment with preferred orientation along the filling 
direction.
As soon as the flow stops, homeotropic domains nucleate in the cell 
and expand until the whole sample becomes homeotropic.
%
\begin{figure}
\begin{center}
\epsfig{file=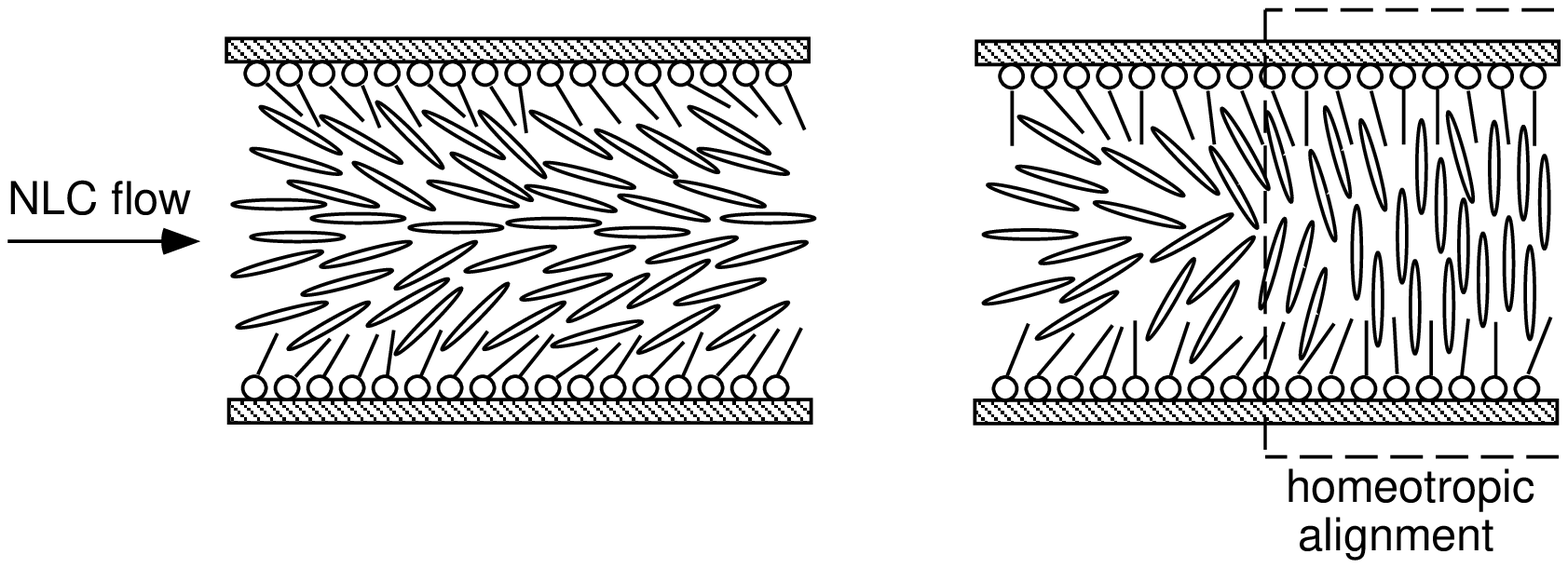, width=0.95\textwidth}
\caption{Left: during filling the LB monolayer is affected by the flow
and the NLC is quasi-planarly aligned with bend-splay deformation.
Right: as soon as the flow ceases the elastic deformation and the LB film 
both contribute to the relaxation to the homeotropic alignment.
\label{relax}
}
\end{center}
\end{figure}
The role of the LB monolayer has been discussed in 
\cite{FazKomLag98a, FazKomLagMot00}: during the filling process the 
chains are distorted by the flow, and relax to the equilibrium position 
once the flow has ceased (Figure \ref{relax}).

\begin{figure}
\begin{center}
\epsfig{file=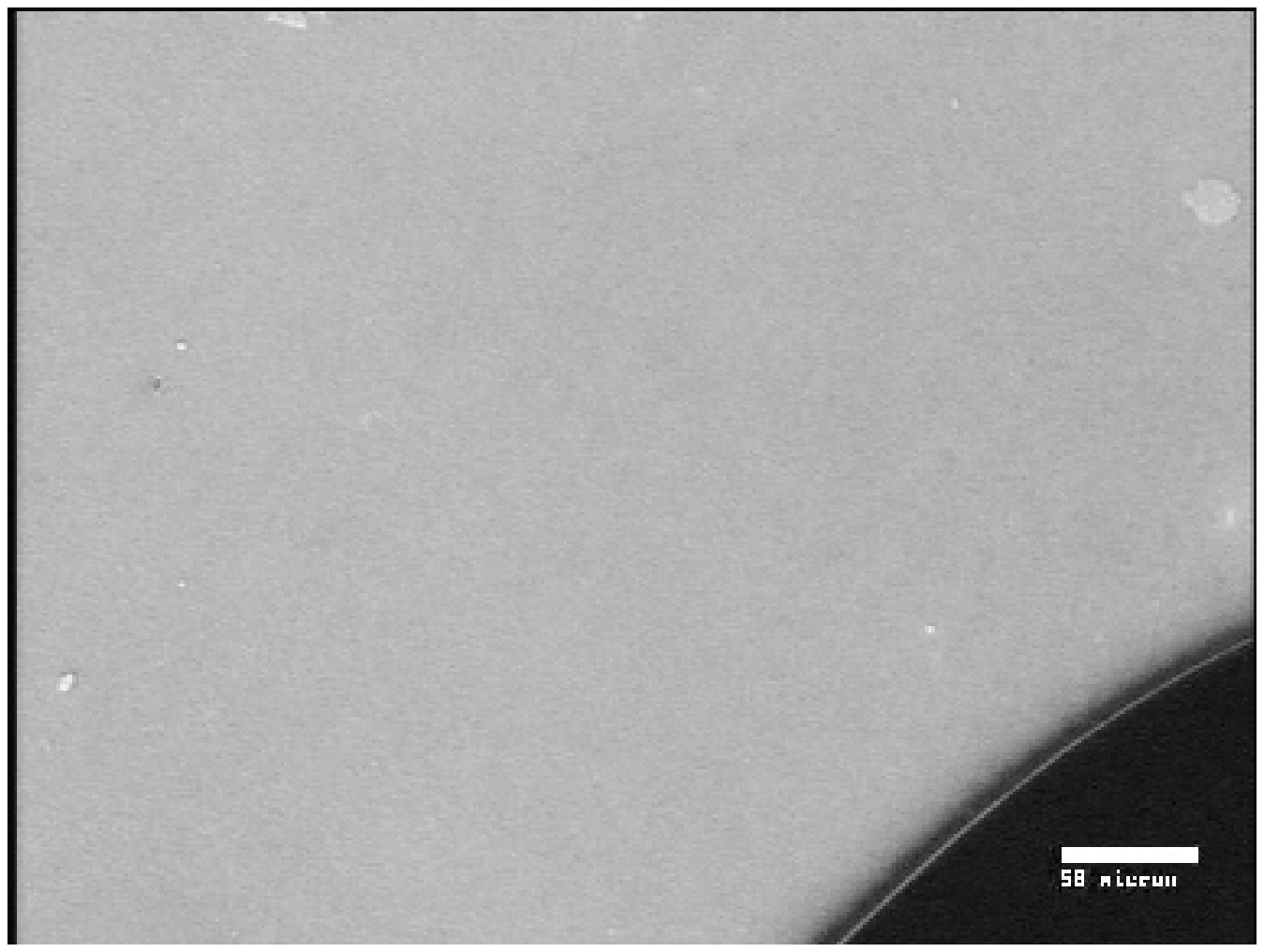, width=0.45\textwidth}
\hspace{0.04\textwidth}
\epsfig{file=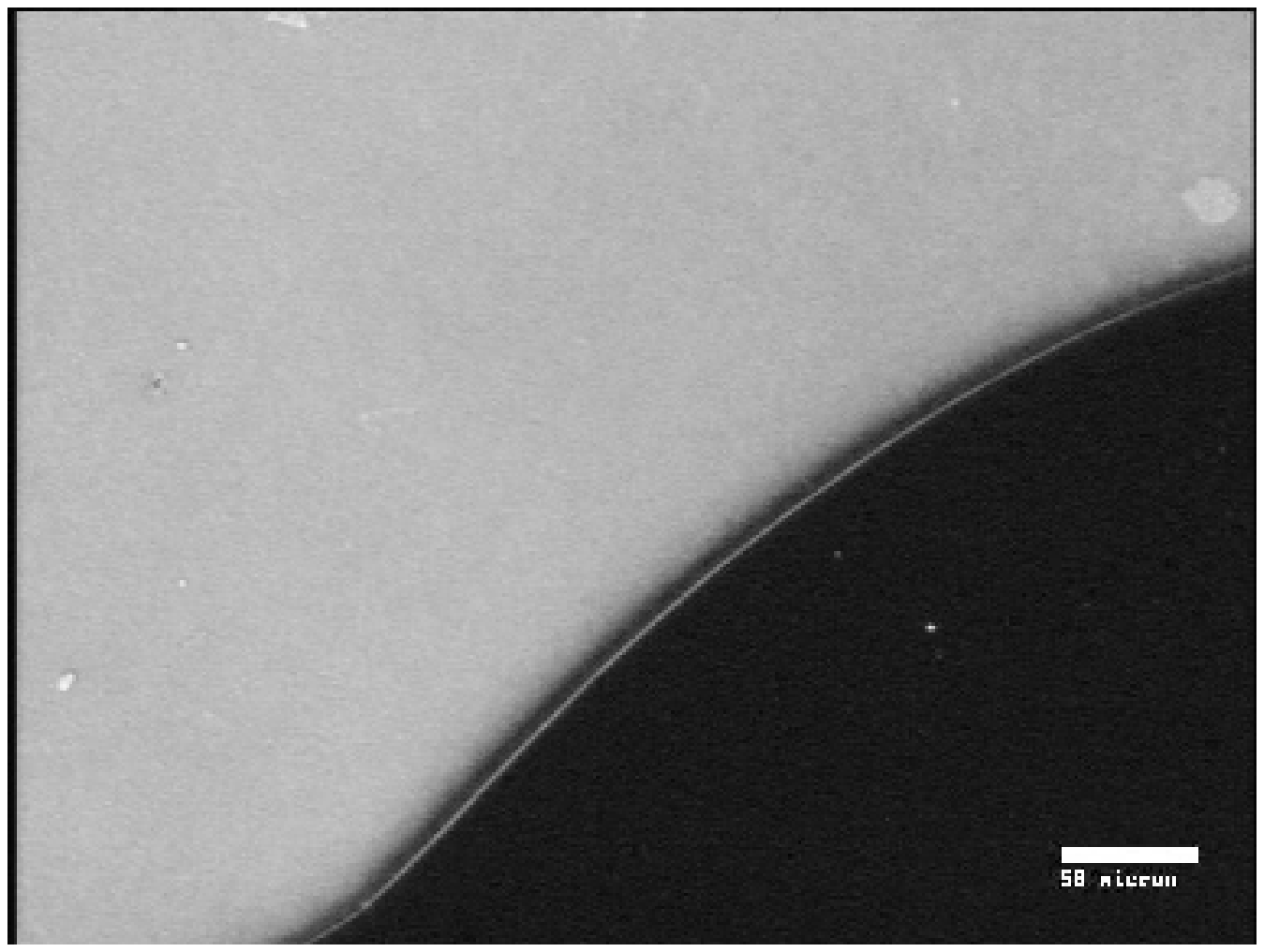, width=0.45\textwidth}
\caption{Cell between crossed polarizers.
A homeotropic domain (dark) expands in the quasi-planar domain.
LB aligning layer: C18; cell thickness: 12.5\,$\mu$m; time interval: 30\,sec. 
\label{relax_cell}
}
\end{center}
\end{figure}

\begin{figure}
\begin{center}
\epsfig{file=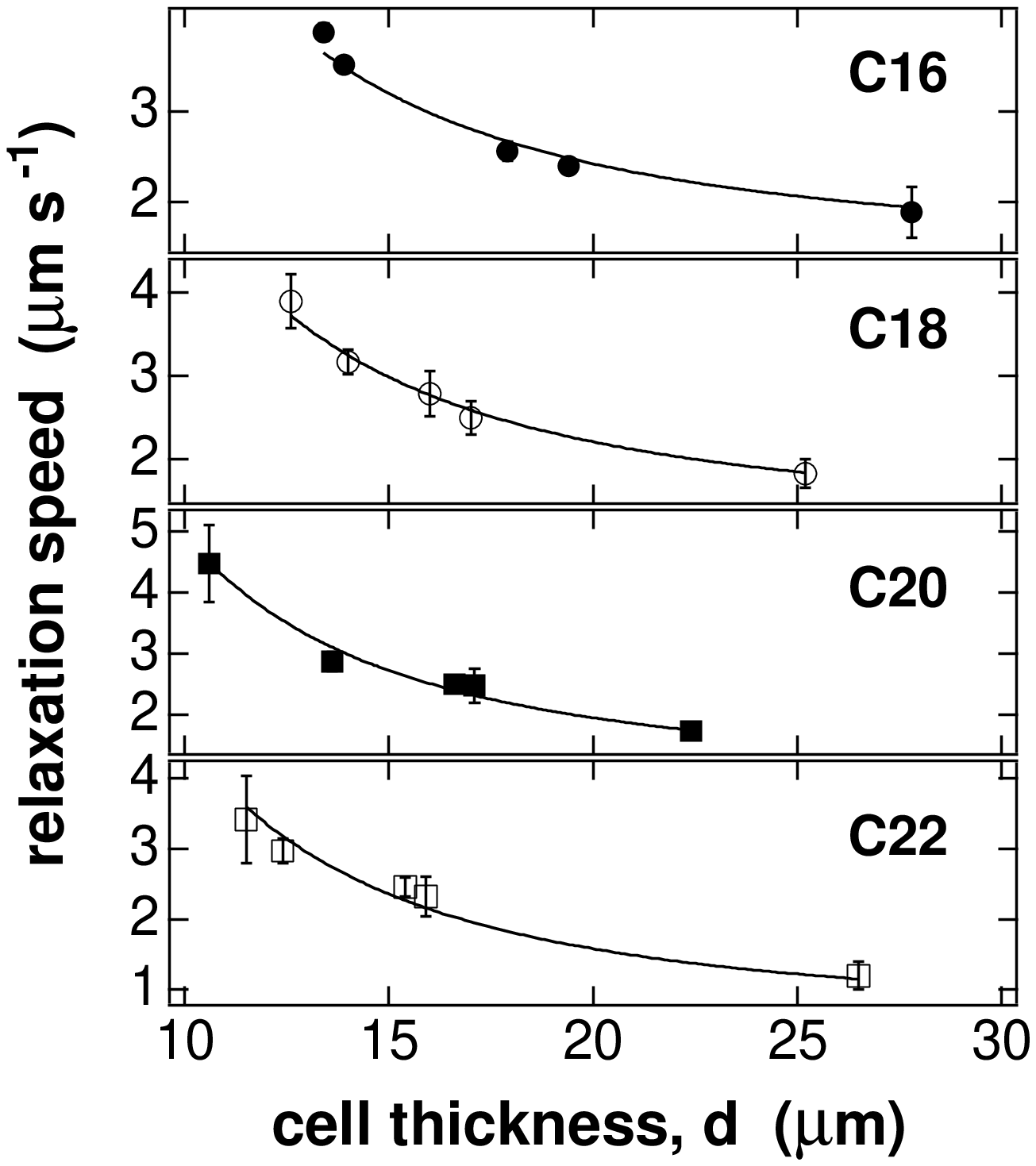, width=0.7\textwidth}
\caption{
\label{speed}
Speed of expansion of the homeotropic domains as a function 
of the cell thickness $d$ for the four LB aligning layers.
The symbols are the experimental values and the lines are fits to the
function in Equation \protect\ref{eqn012}.
The fitting parameter $v_{\text{s}}$ depends on the LB aligning layer,
while $B$ was found do be essencially the same for all LB layers
($B = 400\,\mu\text{m}^{3}$\,s$^{-1} \pm 40\,\mu\text{m}^{3}$m\,s$^{-1}$).
}
\end{center}
\end{figure}

A time sequence of pictures of a homeotropic domain expanding in 
the quasi-planar structure is shown in Figure \ref{relax_cell}.
%
\psfull
The pictures refer to a cell whose plates had been coated with a 
monolayer of C18.
The procedure was repeated for the four aligning fatty acids 
listed in Table \ref{acids} and the speed of expansion of the homeotropic 
domains was found to depend on the length of their alkyl chains and 
on the cell thickness, as shown in Figure \ref{speed}.

The dynamics of the process is governed by surface and bulk effects.
Here we present a simple model in which only two contributions are 
taken into account, namely, a rapid relaxation in the boundary LB 
monolayer (whose characteristic time is $\tau_{\text{s}}$) 
followed by the elastic relaxation of the splay-bend deformation 
in the bulk (whose characteristic time is $\tau_{\text{b}}$).
The total relaxation time, $\tau$, is given by:
\begin{equation}
\frac{1}{\tau} = \frac{1}{\tau_{\text{s}}} + \frac{1}{\tau_{\text{b}}}.
\label{eqn001}
\end{equation}

For a (small) deformation $\theta$ (see Figure \ref{relax})
the surface process is determined by the balance between the anchoring, 
the elastic, and the surface viscosity forces and 
reads\cite{BlinovChigrinov}
\begin{equation}
K \frac{\partial \theta}{\partial z} + w \theta 
= \eta \frac{\partial \theta}{\partial t},
\label{eqn002}
\end{equation}
where $K$ is the characteristic elastic constant of the splay-bend 
deformation in the one-constant approximation, $w$ is the anchoring 
strength, and $\eta$ is the surface viscosity\cite{Pikin}.
The characteristic time $\tau_{\text{s}}$ is given by:
\begin{equation}
\tau_{\text{s}} = \frac{2 w \eta}{w^{2} + K^{2} \beta^{2}},
\label{eqn003}
\end{equation}
where $\beta^{-1}$ is a characteristic length given by
\begin{equation}
\beta^{-2} = \frac{K^{2}- d K w}{w^{2}},
\label{eqn004}
\end{equation}
where $d$ is the cell thickness.
In the case of weak anchoring $K / w d \gg 1$ and thus $\beta^{-1} = 
K/w$, the extrapolation length\cite{Sonin}.
In this case the surface relaxation time becomes  
\begin{equation}
\tau_{\text{s}} = \frac{\eta}{w}.
\label{eqn0041}
\end{equation}

The bulk process is determined by the balance between the elastic 
torque and the bulk viscous torque and reads\cite{BlinovChigrinov}
\begin{equation}
K \frac{\partial^{2} \theta}{\partial z^{2}} =
\gamma \, \frac{\partial \theta}{dt},
\label{eqn005}
\end{equation}
where $\gamma$ is the bulk viscosity.
The characteristic time of the bulk relaxation is 
then\cite{FazKomLag98a}
\begin{equation}
\tau = \frac{\gamma d^{2}}{\pi^{2} K}.
\label{eqn006}
\end{equation}

In our experiment we measured directly the speed with which 
the homeotropic domains expand in the quasi-planar state, i.e.
the speed with which the quasi-planar domain relaxes to homeotropic.
The speed of the surface (bulk) relaxation, $v_{\text{s}}$ ($v_{\text{b}}$) 
is inversionally proportional to the surface (bulk) relaxation time, i.e.
$v_{\text{s}} \propto \tau_{\text{s}}^{-1}$ 
($v_{\text{b}} \propto \tau_{\text{b}}^{-1}$).
Thus, using Equation \ref{eqn001}, the total speed of relaxation 
$v = v_{\text{s}} + v_{\text{b}}$, is given by:
\begin{equation}
v \propto \frac{w}{\eta} + \frac{K \pi^{2}}{\gamma d^{2}}.
\label{eqn007}
\end{equation}

Due to the selective adsorption of ions present in the NLC bulk at 
the surface \cite{BarDur90, AleBarPet93}, $w$ depends on the cell 
thickness.
This ionic charge attracts ions of the opposite charge and a double 
layer of charges is formed which depends only on the substrate and on 
the NLC used.
As a consequence, a surface electric field is created, which extends 
in the bulk over the Debye screening length, $\lambda_{\text{D}}$.
For symmetry reasons the field is normal to the substrate and 
has an orienting effect on the NLC through the coupling to the 
dielectric anisotropy.
The anchoring strength thus appears to be thickness dependent because the 
amount of ions present in the liquid crystal depends on the sample 
volume which is proportional to the cell thickness $d$.
This effective anchoring strength can be written as\cite{AleBarPet93}
\begin{equation}
w = w_{\text{R}} + 
\frac{|\Delta \varepsilon|\,\lambda_{\text{D}}}
{2\,\varepsilon^{2}\,\varepsilon_{0}}\,\sigma^{2},
\label{eqn008}
\end{equation} 
where the flexoelectric effect has been neglected.
In Equation \ref{eqn008} $w_{\text{R}}$ is the Rapini-Papoular anchoring 
strength\cite{RapPap69} which does not depend on the cell thickness.
$\Delta \varepsilon$ is the dielectric anisotropy, $\varepsilon$ is the 
average dielectric constant of the liquid crystal, and 
$\varepsilon_{0}$ is the vacuum dielectric constant. 
$\sigma$ is the electric charge density adsorbed at the surface:
\begin{equation}
\sigma = \Sigma \, \frac{d}{d + 2\,\lambda_{\text{D}}},
\label{eqn009}
\end{equation} 
where $\Sigma$ depends on the conductivity of the liquid crystal and 
on the characteristics of the surface.

Using Equations \ref{eqn008} and \ref{eqn009}, Equation \ref{eqn007} 
becomes:
\begin{equation}
v \propto \frac{w_{\text{R}}}{\eta}
+ \frac{C}{\eta} \left[ {\frac{d}{d + 2\,\lambda_{\text{D}}}}
\right]^{2} 
+ \frac{K \pi^{2}}{\gamma d^{2}},
\label{eqn010}
\end{equation}
where
\begin{equation}
C = \frac{|\Delta \varepsilon|\,\lambda_{\text{D}}}
{2\,\varepsilon^{2}\,\varepsilon_{0}} \Sigma^{2}
\label{eqn011}
\end{equation}
is a constant characteristic of the liquid crystal and the substrate.
Generally $\lambda_{\text{D}} \ll d$\cite{BarDur90}
and the second term of the 
right-hand side of Equation \ref{eqn010} varies very little with $d$ as 
compared to the third term.
Thus, the \textit{surface speed} $v_{\text{s}}$ can be considered 
essentially constant with respect to the \textit{bulk speed} 
$v_{\text{b}}$ in the range of cell thickness used in this work.
Equation \ref{eqn010} can thus be rewritten as
\begin{equation}
v = v_{\text{s}} + B d^{-2},
\label{eqn012}
\end{equation}
where $v_{\text{s}} = (w_{\text{R}} + C )/\eta = $ constant and 
$B$ is a proportionality factor which depends on $K$ and $\gamma$.

\begin{figure}
\begin{center}
\epsfig{file=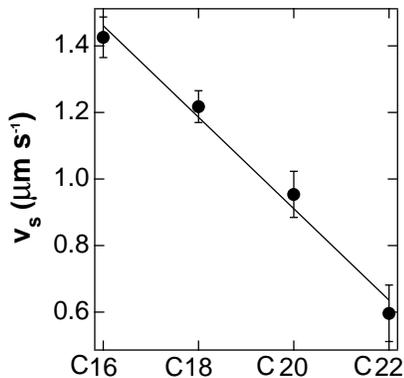, width=0.45\textwidth}
\caption{
\label{v_s}
\textit{Surface-speed} of expansion of the homeotropic domains as a 
function of the aligning monolayer chain length.
The agreement with a linear decrease of $v_{s}$ with increasing the 
length of the chain is very good.}
\end{center}
\end{figure}

The data in Figure \ref{speed} where fitted to the function in 
Equation \ref{eqn012} and the fits are shown in the figure.
The fitting parameter $v_{\text{s}}$ was found to depend on the LB 
aligning monolayer, as shown in Figure \ref{v_s}, which is in 
agreement with our interpretation of it as characteristic of the surface.
The fitting parameter $B$ was found to be essentially constant in the 
error bar ($B=400\,\mu\text{m}^{3}\,$s$^{-1} \pm 
40\,\mu\text{m}^{3}$\,s$^{-1}$) independent of the aligning monolayers. 
The surface relaxation is faster for shorter chains, and seems to follow a 
linear relation with the number of carbon atoms in the chain.
Moreover, the quality of the homeotropic alignment itself is higher for 
shorter chains.
These two observations together suggest that the anchoring of MBBA is 
quite sensitive even to small differences in the length of the 
surfactant.
For instance, the anchoring strength is expected to be affected by the 
presence of chains of different length in the same monolayer.

\begin{figure}
\begin{center}
\epsfig{file=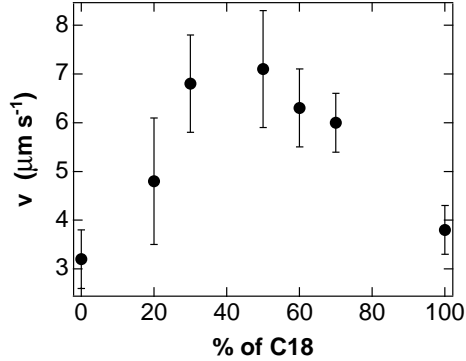, width=0.5\textwidth}
\caption{
\label{vC18C22mix}
Speed of expansion of the homeotropic domains as a 
function of the percentage of C18 in the mixed C18:C22 
aligning LB monolayer. 
Cell thickness 15\,$\mu$m.}
\end{center}
\end{figure}

The relaxation process was also studied in cells of fixed thickness (15\,$\mu$m) 
coated with mixed LB films of C18 and C22 in different ratios.
Figure \ref{vC18C22mix} shows the speed of relaxation of the homeotropic 
domains as a function of the percentage of C18 in the LB aligning 
monolayer.
The relaxation is faster for the mixed layers in which the 
proportions of the two compounds are similar.
In particular, the highest speed is obtained for the 50:50 mixture.
Also in this case the quality of the homeotropic alignment was higher 
in those cells where the relaxation process was faster.

\subsection{Anchoring strength}
\begin{figure}
\begin{center}
\epsfig{file=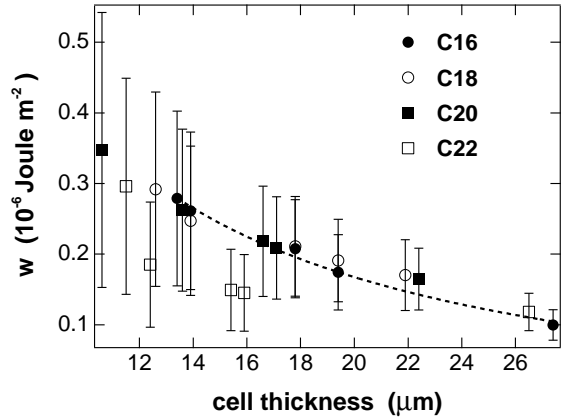, width=0.6\textwidth}
\caption{
\label{wC16C18C20C22}
Anchoring strength of MBBA on the four fatty acid LB monolayers.
The anchoring is weak\protect\cite{FazKom99} and the rather large error 
of the method does not allow to resolve small differences.
}
\end{center}
\end{figure}

The anchoring strength MBBA on the fatty acid LB monolayers has been measured 
with the Frederiks transition method\cite{Sonin}.
MBBA possesses negative dielectric anisotropy 
and thus a vertical electric
field destabilizes MBBA's homeotropic alignment.
Homeotropic alignment appears dark between crossed polarisers.
Above a certain electric field threshold (Frederiks transition threshold) 
the molecules in the bulk are tilted by the electric field and some light 
is transmitted through the crossed polarisers.
Here we defined as threshold field $E_{\text{t}}$ the one at which 20{\%} 
of the saturated value of the transmitted light intensity was reached.
The anchoring strength $w$ can then be calculated 
as\cite{FazKom99, FazioValentina00}:
\begin{equation}
w = q\,K\,\tan\left( \frac{q\,L}{2} \right),
\,\,\,\,\,\,\,\,\,\,
\text{with}
\,\,\,\,\,\,\,\,\,\,
q = E_{\text{t}} \sqrt{ \frac{|\Delta 
\varepsilon|\,\varepsilon_{0}}{K} }.
\label{eqn03}
\end{equation} 
In Figure \ref{wC16C18C20C22} the dependence of $w$ on the cell 
thickness is presented.
Because of the weak anchoring the small differences in anchoring 
strength due to the different aligning layers cannot be resolved 
using this method.

\begin{figure}
\begin{center}
\epsfig{file=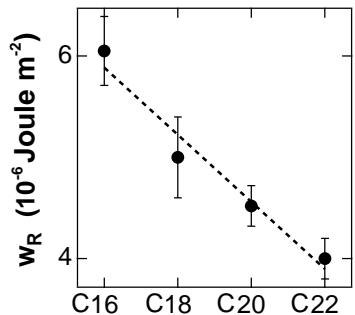,width=0.4\textwidth}
\caption{
\label{wRC16C18C20C22}
Rapini-Papoular anchoring strength of MBBA on fatty acids.
The trend is similar to that of the relaxation speed of the 
homeotropic alignment: the agreement with a linear decrease of $w_{R}$ 
with increasing the length of the chain is rather good.
}
\end{center}
\end{figure}

The data in Figure \ref{wC16C18C20C22} are fitted very well by 
Equation \ref{eqn008} (in Figure \ref{wC16C18C20C22}
only the fit for the C16 data is shown).
$w_{\text{R}}$ and $\Sigma$ are the fitting parameters. 
$\Sigma$ was found to be $(1.3 \pm 0.3) \times 10^{-4}$ C\,m$^{-2}$
independent on the cell thickness.
Instead, $w_{\text{R}}$  was found to decrease with increasing the 
chain length of the surfactants, as shown in Figure \ref{wRC16C18C20C22}.
This dependence reminds that of the relaxation speed of the homeotropic 
alignment (Figure \ref{v_s}): the agreement with a linear decrease of 
$w_{\text{R}}$ with increasing the length of the chain is rather good.

\subsection{Anchoring breaking}
\begin{figure}
\begin{center}
\epsfig{file=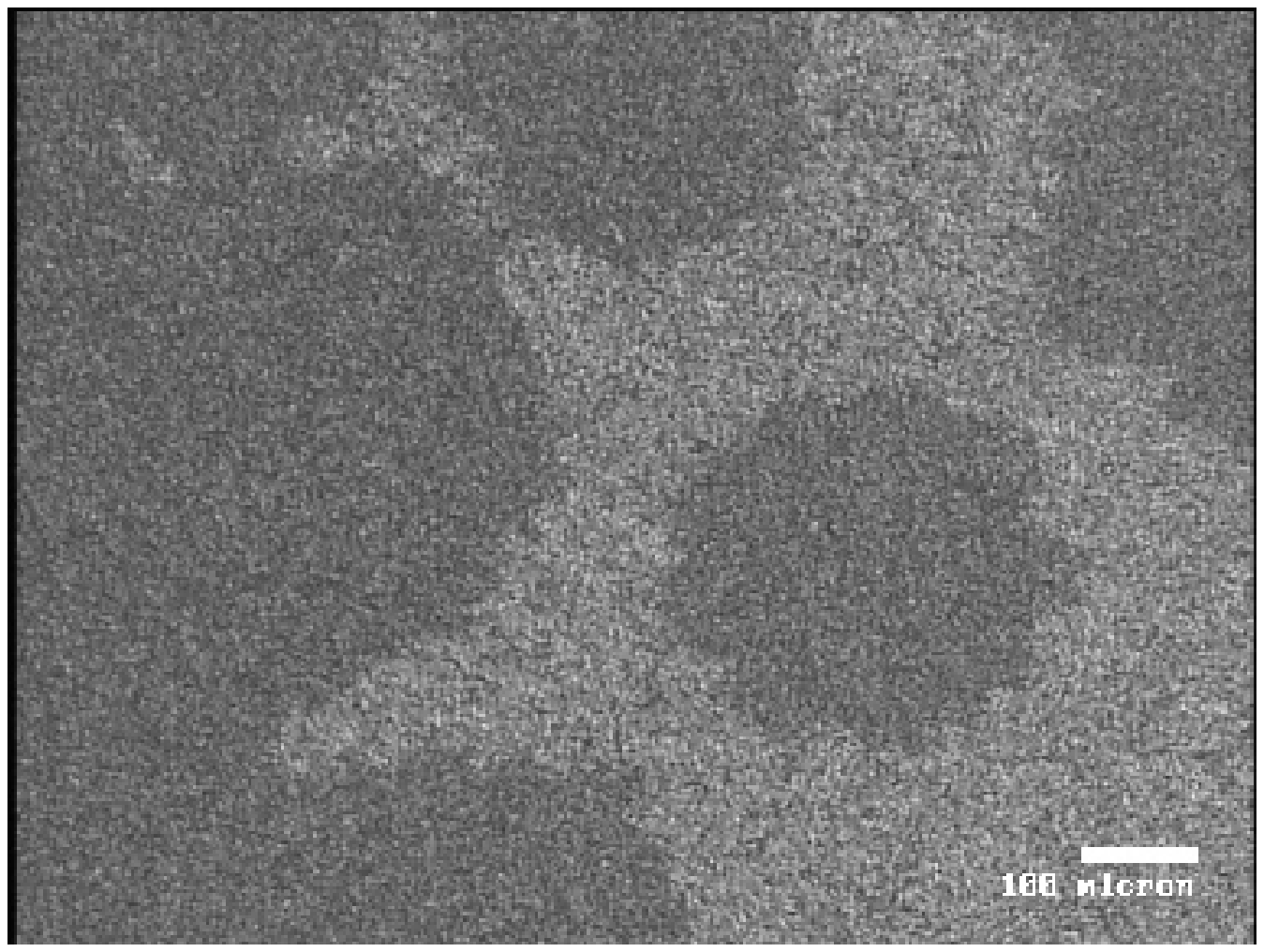, width=0.45\textwidth}
\hspace{0.04\textwidth}
\epsfig{file=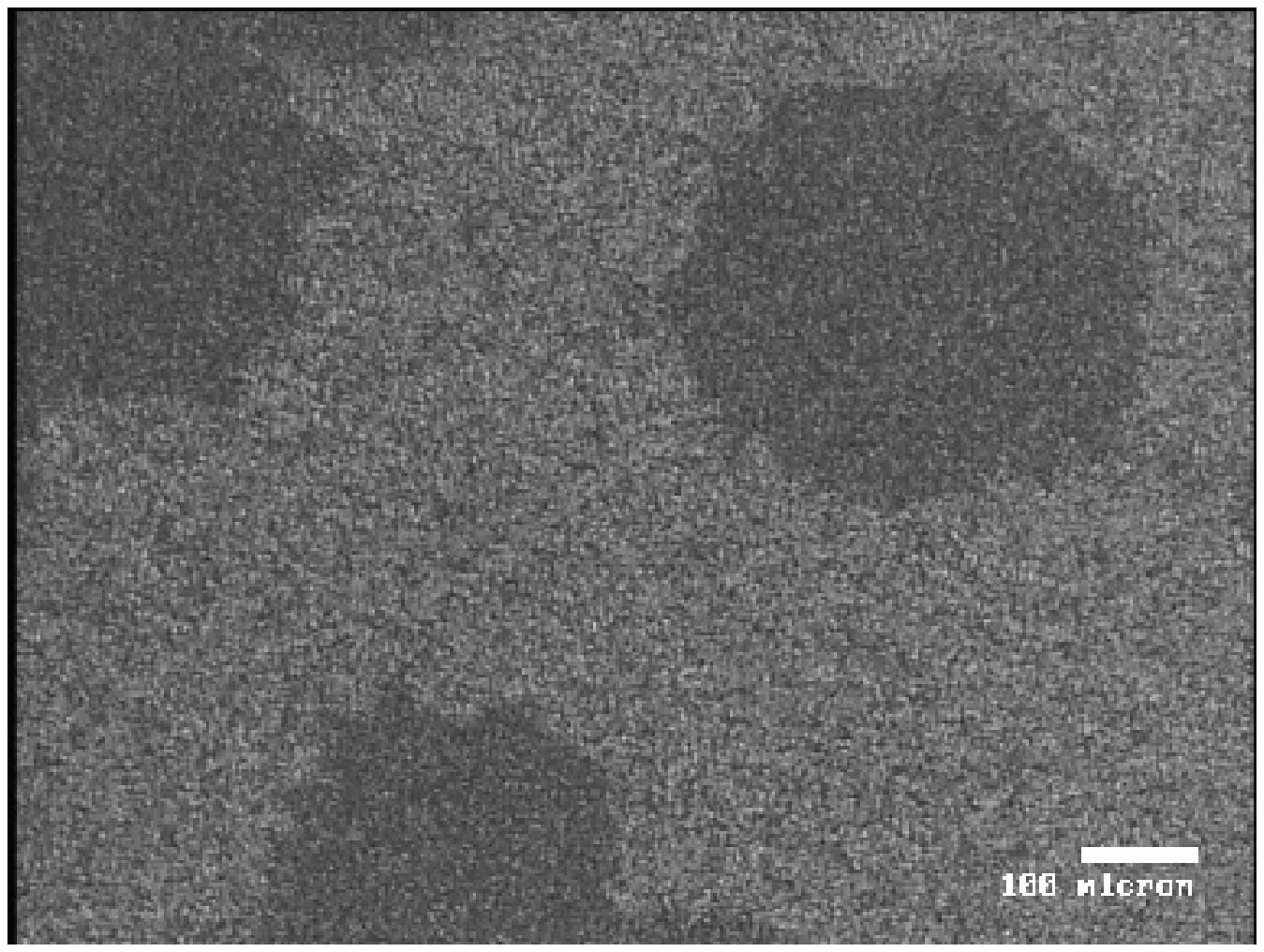, width=0.45\textwidth}
\caption{
\label{DSM2fig}
DSM2 domains expand in the DSM1 turbulent state.
Note the circular form of the DSM2 domains, due to absence of a 
preferred alignment direction in the plane of the cell because of the 
initial homeotropic alignment\protect\cite{FazKom99,LucScaStrVer99}
(in planar cells they expand faster in the rubbing direction 
\protect\cite{ScaVerCar95,StrVerScaCar99}).
}
\end{center}
\end{figure}

Recently\cite{FazKom99} we have shown that an alternative method for 
measuring/comparing anchoring strengths is the mechanism of anchoring 
breaking due to the onset of an electric-field--induced 
turbulence-to-turbulence transition in the sample.
Above the Fredericks transition threshold initially homeotropic MBBA
(aligned by a fatty acid monolayer) is essentially quasi-planarly oriented.
Increasing the field above the Fredericks threshold one observes first 
modulate structures (William rolls \cite{BlinovChigrinov}), then their 
destabilization, and then the transition to a weakly turbulent state, 
called DSM1 (Dynamic Scattering Mode 1).
All these electric-field--induced instabilities do not affect the 
anchoring of MBBA on the fatty acid monolayer,
and the homeotropic alignment is immediately restored 
once the electric field is switched off \cite{FazKom99}.
Increasing the electric field further above the DSM1 threshold, one 
observes the transition to another turbulent state, denoted as DSM2:
domains of DSM2 nucleate in the nematic layer which is in the DSM1 
state, and expand over the whole sample area in form of circular 
domains\cite{FazKom99, LucScaStrVer99} (Figure \ref{DSM2fig}).
%
\psfull
In \cite{FazKom99} we demonstrated that DSM1-to-DSM2 transition is 
due to the breaking of the surface anchoring: switching off the 
electric field above the DSM2 transition, the homeotropic alignment is 
not immediately restored.
Instead, the sample is in a metastable quasi-planar state which 
relaxes to homeotropic as the flow-induced quasi-planar state does 
(in form of circular homeotropic domains that expand until the whole sample 
becomes homeotropic again).
Thus, the electric field threshold of DSM2 is proportional to the 
anchoring strength and can be considered as a measure of it.

\begin{figure}
\begin{center}
\epsfig{file=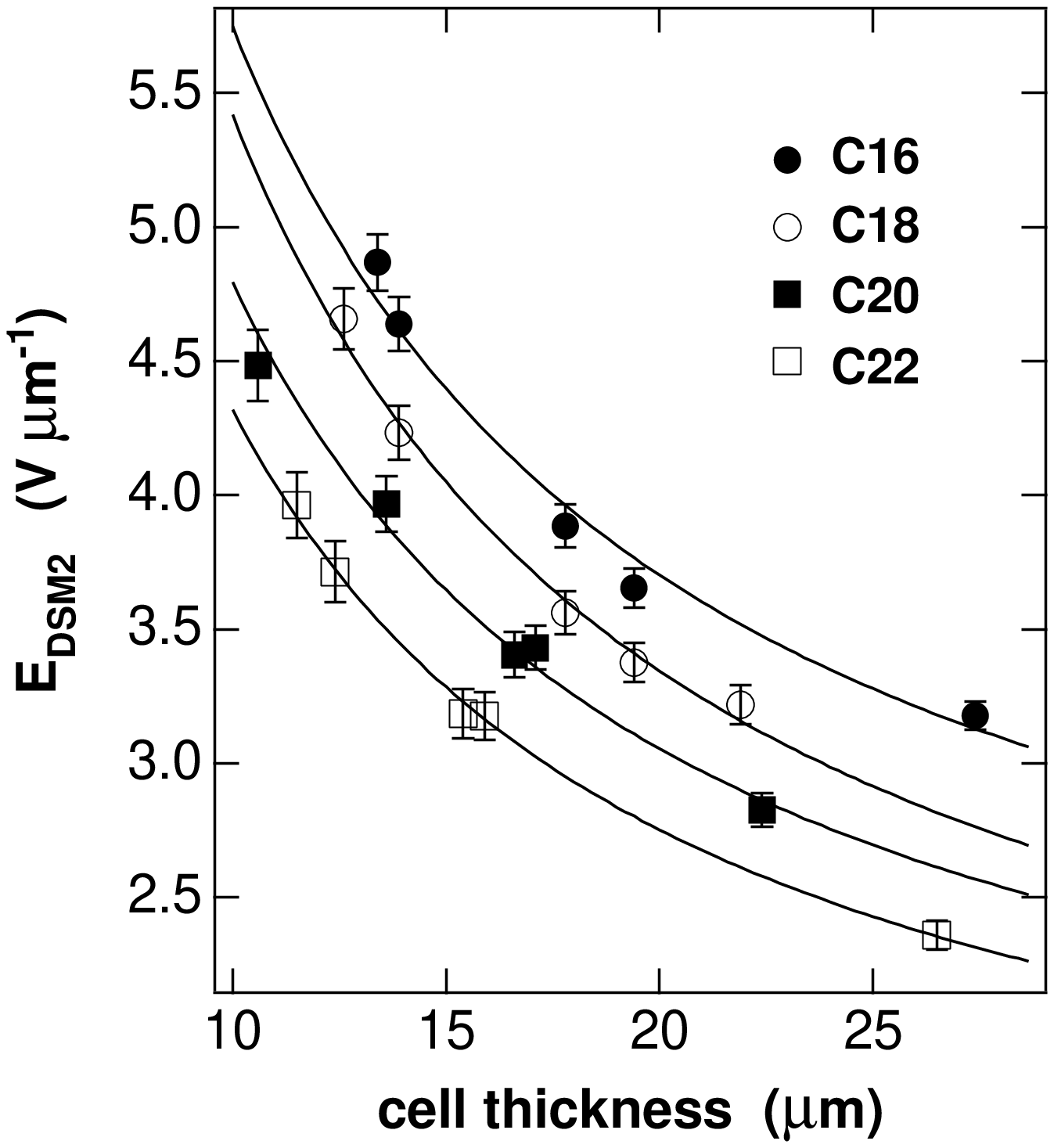, width=0.8\textwidth}
\caption{
\label{DSM2thres}
Threshold for the onset of the DSM2 turbulent state for MBBA on fatty 
acid LB monolayers as a function of the cell thickness.
$E_{\text{DSM2}}$ depends on the length of the alkyl chain of the 
fatty acids: the differences in anchoring strength of MBBA on the 
different LB monolayers are therefore resolved.
The lines are fits to the function in Equation \protect\ref{eqn008}.
}
\end{center}
\end{figure}

\begin{figure}
\begin{center}
\epsfig{file=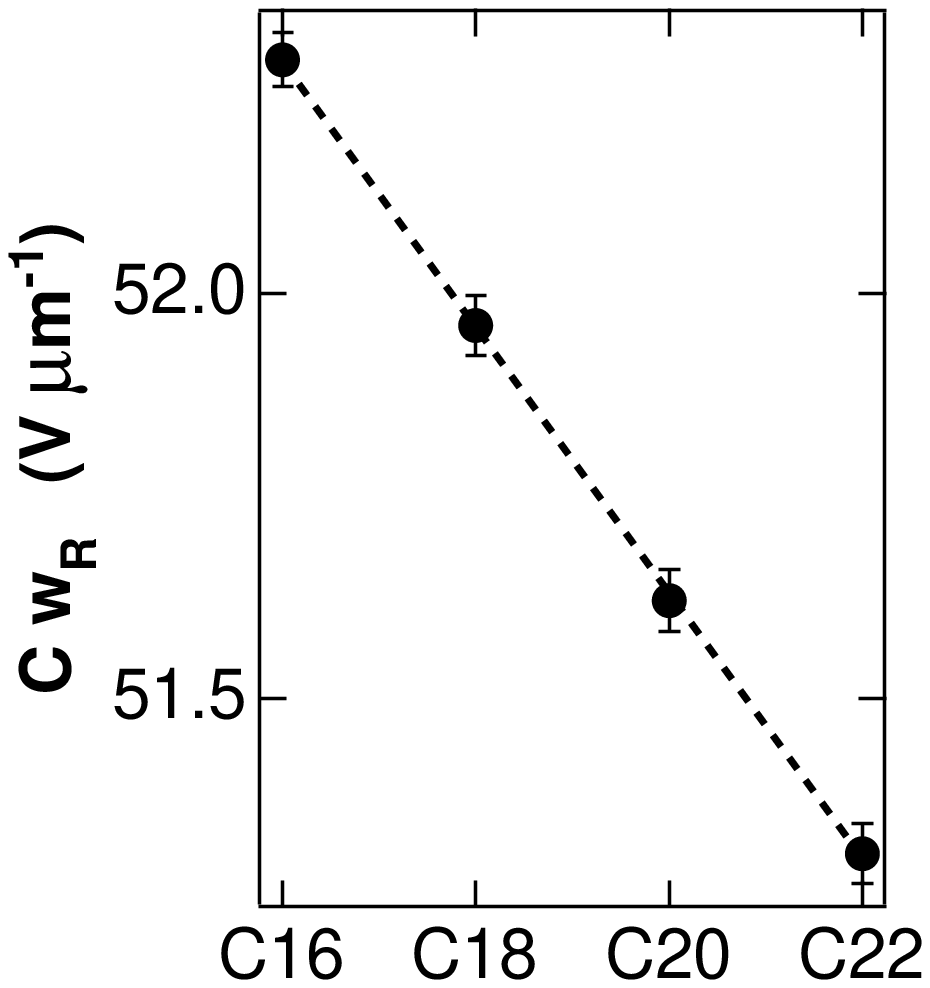, width=0.4\textwidth}
\caption{
\label{Cw_R}
The fitting parameter $C w_{\text{R}}$ is proportional to $w_{\text{R}}$
and thus decreases linearly with increasing the length of the alkyl chains.
}
\end{center}
\end{figure}

In Figure \ref{DSM2thres} the electric field threshold
for DSM2, $E_{\text{DSM2}}$, in MBBA cells are 
shown as a function of the cell thickness for the four fatty acid LB 
aligning layers.
There is a clear dependence on the nature of LB monolayer which demonstrates 
how sensitive this method is in resolving the small differences arising from 
the different thickness of the aligning monolayers.
Since the threshold for DSM2 is proportional to the anchoring 
strength, the data were fitted with a function proportional to the 
anchoring strength in Equation \ref{eqn008}:
\begin{equation}
f = C \left( w_{\text{R}} + 
\frac{|\Delta \varepsilon|\,\lambda_{\text{D}}}
{2\,\varepsilon^{2}\,\varepsilon_{0}}\,\sigma^{2} \right),
\label{eqn040}
\end{equation}
where $C$ is a constant of proportionality.
The fitting parameter $C w_{\text{R}}$ is proportional to $w_{\text{R}}$
and thus decreases linearly with increasing the length of the alkyl chains 
(Figure \ref{Cw_R}) as $w_R$ (Figure \ref{wRC16C18C20C22}). 

\begin{figure}
\begin{center}
\epsfig{file=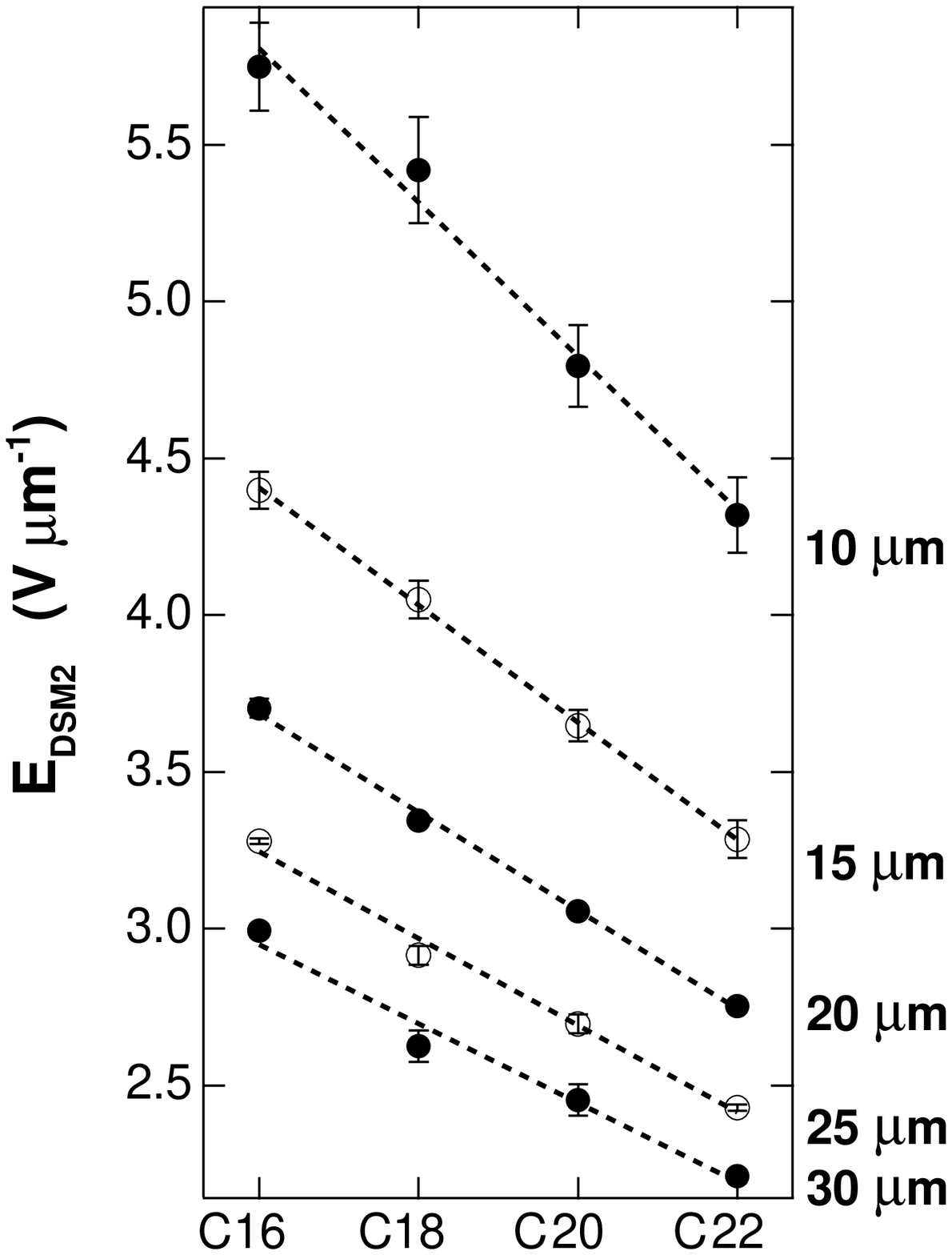, width=0.5\textwidth}
\caption{
\label{DSM2chain}
Threshold for the onset of the DSM2 turbulent state for MBBA on fatty 
acid LB monolayers as a function of the length of the fatty acid 
alkyl chains, for different cell thickness.
$E_{\text{DSM2}}$ decreases linearly with increasing the chain length.
}
\end{center}
\end{figure}

From the fits it is possible to extrapolate the value of $E_{\text{DSM2}}$ 
as a function of the LB chain length for different cell thicknesses, which 
are plotted in Figure \ref{DSM2chain}.
$E_{\text{DSM2}}$ is linearly decreasing with increasing the chain length,
as $v_{\text{s}}$ and $w_{\text{R}}$.

\section{Discussion and conclusions}
Whereas the present knowledge about alignment of liquid 
crystals (especially nematics) may be considered as sufficient for 
achieving a desired kind of alignment, the tuning of the anchoring 
strength is a very difficult task, if not impossible.
The difficulties in controlling the anchoring strength arise not only 
from the lack of proper aligning materials, but also from the lack of 
sensitive enough methods to resolve small variations in anchoring 
strength.

In this work, two methods for estimating the anchoring strength of  
homeotropically aligned NLCs on LB monolayers of fatty acids have been 
presented.
One of them is field-free, whereas in the other one an external electric 
field is necessary.

In the field-assisted method, the threshold for the onset of the 
DSM2 turbulent state can be interpreted as a measure of the 
anchoring strength, because above that threshold the anchoring of the 
liquid crystal to the substrate is broken.
For a fixed cell thickness $E_{\text{DSM2}}$ decreases linearly with 
increasing the length of the surfactant chains.
The \lq\lq surface speed\rq\rq\, of relaxation from flow-induced 
quasi-planar to surface-induced homeotropic alignment also decreases 
linearly with increasing the length of the surfactant chains, and can 
thus as well be interpreted as a measure of the anchoring strength. 
The two methods not only present a quite high sensitivity, as they can 
resolve very small differences in anchoring strength due to the use of 
slightly different surfactants, but give also independent and consistent
results.

\begin{figure}
\begin{center}
\epsfig{file=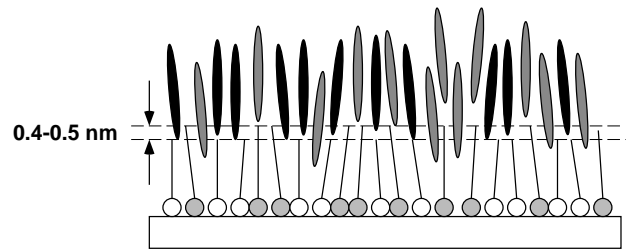, width=0.6\textwidth}
\caption{
\label{penetration}
Alignment of MBBA on mixed LB monolayers of C18 (white) and C22 (gray).
Part of the LC molecules (gray) penetrate into the region of the 
surfactant, but part of them (black) are aligned because of the 
difference in the length of the alkyl chains, about 0.4-0.5\,nm 
(interdigitation).
}
\end{center}
\end{figure}

In cells treated with mixed C18:C22 LB monolayers it was found that 
the relaxation speed, and thus the anchoring strength, strongly depend
on the mixture used and it is correlated with the quality of the 
homeotropic anchoring.
In particular the relaxation speed is  
maximum for similar concentrations of the two fatty 
acids in the mixed monolayer.
In this case, in fact, not only the LC is aligned by penetration into 
the surfactant chain region, but also by the penetration into the \lq\lq 
holes\rq\rq\, created by the differences in chain lengths 
(interdigitation).
A schematic picture is shown in Figure \ref{penetration}.
This suggests a way of \lq\lq tuning\rq\rq\, the anchoring strength: 
$w$ can be changed by varying the ratio of the components of the LB film, 
and thus the density of \lq\lq holes\rq\rq\,.
That the LB film density influences the anchoring of nematics is 
known \cite{AleBarBarBon95}, but the phenomenon of interdigitation 
has yet not been addressed.

In conclusion, the methods presented in this work make it possible to 
carry out detailed studies on the anchoring of nematic liquid crystals 
on surfactants.
Their sensitivity allows the use of a great variety of surfactants or, 
more generally, materials that induce homeotropic alignment.
On the basis of the obtained results new theoretical models can be 
developed in order to reach a deeper understanding of the 
microscopic LC-surfactant interactions.

\section{Acknowledgments}
Valentina S. U. Fazio acknowledges the TMR European programme (contract 
number ERBFMNICT983023) for financial support.
Francesca Nannelli acknowledges the Lerici Foundation for financial 
support.
\bibliography{journal2,a2.bib}
\end{document}